# Perspectives in and On Quantum Theory

Richard Healey,

Emeritus Professor of Philosophy,

University of Arizona,

Tucson, Arizona, USA

Email: richardahealey@gmail.com

orcid=0000-0001-9329-2927

## Abstract

I take a pragmatist perspective on quantum theory. This is not a view of the world described by quantum theory. In this view quantum theory itself does not describe the physical world (nor our observations, experiences or opinions of it). Instead, the theory offers reliable advice—on when to expect an event of one kind or another, and on how strongly to expect each possible outcome of that event. The event's actual outcome is a perspectival fact—a fact relative to a physical context of assessment. Measurement outcomes and quantum states are both perspectival. By noticing that each must be relativized to an appropriate physical context one can resolve the measurement problem and the problem of nonlocal action. But if the outcome of a quantum measurement is not an absolute fact, then why should the statistics of such outcomes give us any objective reason to accept quantum theory? One can describe extensions of the scenario of Wigner's friend in which a statement expressing the outcome of a quantum measurement would be true relative to one such context but not relative to another. However, physical conditions in our world prevent us from realizing such scenarios. Since the outcome of every actual quantum measurement is certified at what is essentially a single context of assessment, the outcome relative to that context is an objective fact in the only sense that matters for science. We should accept quantum theory because the statistics these outcomes display are just those it leads us to expect.

**Keywords** Perspectives . Quantum theory . Pragmatist view . Quantum states . Measurement outcomes . Objectivity

## Statements and Declarations

**Competing Interests** While no funds directly supported the research leading to these results, this article was written while I was entitled to receive travel funding from the John Templeton Foundation (under Grant IDs 62424, 62312): the opinions expressed in this work are those of the author and do not necessarily reflect the views of the John Templeton Foundation.

**Funding Declaration**

No funds directly supported the research leading to these results.

# 1. Introduction

A hundred years after the ground-breaking work of Heisenberg, Schrödinger and others, we still don't agree on how to understand the enormously successful theory that grew from the roots they planted. By taking a pragmatist approach I have developed a view of quantum theory and shown how it avoids problems that plague rival views. But I find the continued existence of so many radically different interpretations of quantum theory both extraordinary and disconcerting insofar as it threatens to undermine the objectivity of the scientific knowledge that comes with acceptance of that theory. The first step to securing the necessary objectivity is to understand how quantum theory can be so successfully applied without itself saying what the physical world is like, in accordance with the pragmatist precept that meaning derives from use, not representation. Section 2 suggests that it is best to take this step by rejecting an influential but overly narrow conception of interpretation, replacing it by a new perspective on how to understand quantum theory.

From this pragmatist perspective, key elements of quantum theory—quantum states and measurement outcomes—are themselves perspectival. By noticing that each must be relativized to an appropriate physical context one can resolve the measurement problem as well as the problem of nonlocal action. But this raises a new problem for the objectivity of knowledge based on acceptance of quantum theory. If the outcome of a quantum measurement is not an absolute fact, then why should the statistics of such outcomes give us any objective reason to accept quantum theory? The rest of the article seeks to clarify and answer this question.

From section 3 I adopt a pragmatist perspective on quantum theory, relying on the results of my published research papers collected in [1]. (Readers seeking to consult these papers prior to publication of [1] will find detailed references to eight published before 2017 in the Acknowledgements to my introductory book [2].) This is not a perspective on the physical world according to quantum theory: the theory does not say what the quantum world is like, and a quantum state does not describe any intrinsic physical property of a system to which it is assigned. From this viewpoint, quantum theory does not itself describe or represent the physical world (nor our observations, experiences or opinions of it). Instead, the theory offers good advice—on when to expect an event of one kind or another, and on how strongly to expect each possible outcome of that event.

Section 4 explains the sense in which such an event's actual outcome is a perspectival fact—a fact relative to a physical context of assessment. Such contexts are simply natural occurrences: they do not involve the presence, or even the existence, of any observer or agent making the assessment. One can describe extensions of the scenario of Wigner's friend (EWFSs) in which a statement expressing the outcome of a quantum measurement would be true relative to one such context but not true relative to another. The outcome, if any, would be relative to whichever context a scientist considered. However, physical conditions in our world prevent us from realizing such a scenario. Since the outcome of every actual quantum measurement is certified at what is essentially a single context of assessment, the outcome relative to that context is an objective fact in the only sense that matters for science. The statistics of such objective outcomes constitute an overwhelming reason to accept quantum theory.

## 2. Perspectives, Viewpoints and Relativism

To quote the entry on Perspectivism in Science from the *Internet Encyclopedia of Philosophy* [3],

> Perspectivism, or perspectival realism, has been discussed in philosophy for many centuries, but as a view about science, it is a twenty-first-century topic. Although it has taken many forms and even though there is no agreed definition, perspectivism at its heart uses a visual metaphor to help us understand the scope and character of scientific knowledge.

The literal basis of the metaphor is how things look from a particular point of view—the physical location and orientation of a normal human observer. One can abstract from this basis in various ways. One can take the perspective to stay the same even if the observer is not a normal human, and even though no observer then occupies this viewpoint. One can take a viewpoint to include the time as well as the location from which things look that way, as well as the state of the environment at that time (lighting, weather, etc.).

The notion of a viewpoint is sufficiently flexible that one can further abstract by building into it the beliefs of the historically and socially situated observer. With the final abstraction of the notion of observation to include any kind of epistemic access, one arrives at the idea of a perspective on a body of knowledge, scientific or otherwise. This is Wikipedia's formulation of perspectivism [4]:

> Perspectivism is the epistemological principle that perception of and knowledge of something are always bound to the interpretive perspectives of those observing it.

Quantum theory is a fundamental part of contemporary scientific knowledge. But physicists and philosophers seeking to understand this theory have adopted very different interpretive perspectives on it. In a 2016 preprint entitled "Interpretations of quantum theory: A map of madness" [5], Adan Cabello quoted David Mermin [6] as saying that "quantum theory is the most useful and powerful theory physicists have ever devised. Yet today [then] nearly 90 years after its formulation, disagreement about the meaning of the theory is stronger than ever. New interpretations appear every day. None ever disappear." Cabello himself continued "This situation is odd and is arguably an obstacle for scientific progress, or at least for a certain kind of scientific progress."

It is not a problem that an elephant looks different from different viewpoints (from one or other side, from the front, from the back, from on top, from underneath, …). What obstacle to scientific progress is created by the proliferation of alternative interpretations of the theory? Mermin rightly notes that proponents of different interpretations of quantum theory disagree about how the theory should be understood—they treat their interpretations as rivals, at most one of which could be correct. In the ancient parable of the elephant, that is how the blind men treated their different beliefs about the elephant each examined before coming to realize that they all had different tactile perspectives on a single large animal.

This illustrates a general issue faced by perspectivism. Is it possible to say in absolute (non-perspectival) terms what the various perspectives are perspectives on? If it is (it's an elephant!) then perspectival talk is secondary or even dispensable, and perspectivism collapses into a more traditional realism. If it is not, then isn't perspectivism simply relativism by another name? To answer these questions when the perspectives are interpretations of quantum theory, I need to say what I mean by 'quantum theory' and 'interpretation'.

As David Wallace [7] has pointed out, quantum theory is a framework theory, under which many specific quantum theories stand. We have non-relativistic quantum mechanics, relativistic quantum mechanics, specific quantum field theories (both relativistic and non-relativistic), tentative quantum theories of gravity including string theory, and so on. While these theories all have what Wittgenstein [8] called a family resemblance, they don't all make the same assumptions about the structure of space-time, and many are often taken to differ in their ontologies (fields or particles)? Nevertheless, they do share some abstract structural features: all their models contain some mathematical object (such as a wave-function) whose application generates probabilities, and other mathematical objects (linear operators) associated with measurable magnitudes (as the operator $-i\hbar\partial/\partial x$ is associated with *x*-component of momentum in the position-representation of non-relativistic quantum mechanics). A proposed interpretation may apply only to specific quantum theories, or it may aspire to apply to every quantum theory in the framework.

But what is an interpretation of a theory? Bas van Fraassen [9] gave this influential answer:

> An interpretation of a theory is an answer to the question 'How could the world possibly be how this theory says it is?' .

I shall call this notion of theory interpretation interpretation$_{vF}$ Note two things about this answer to the question. It assumes that to understand the meaning of a theory is to know what the world could be like according to that theory. But it allows that one may increase one's understanding by allowing for multiple, incompatible answers, each interpretation offering its own different account of what the world is like. As a constructive empiricist, van Fraassen maintains that a scientist's acceptance of a scientific theory does not require belief in what that theory says about unobservable matters. Consequently, by accepting quantum theory a scientist is not committed to believing what the unobservable world is like according to any particular interpretation$_{vF}$.

If van Fraassen were right, it might seem that disputes among scientists arising from their rival interpretations of quantum theory need have no effect on their scientific work, and so no impact on the progress of science. But even he admits that acceptance of a theory involves more than belief in its empirical consequences—it further involves immersion in the theory to the extent of using it to answer scientific questions and basing one's research on it. So even a constructive empiricist must take a dispute about the unobservable structures figuring in a theory's models to impact the progress of science. A traditional realist should reach the same conclusion more directly: for her, a dispute about the interpretation$_{vF}$ of a quantum theory may well present an obstacle to scientific progress because until that dispute is resolved there will be no general agreement on the nature of the world that quantum theory describes. But if quantum theory does not describe the world, then this lack of agreement is no bar to progress through its successful application.

## 3. A pragmatist perspective that is not an interpretation_vF.

In my pragmatist view, (a) quantum theory does not say how the world is, and so does not admit of any interpretation_vF. What I offer is not yet another interpretation of quantum theory but a new perspective on quantum theory that seeks understanding not by asking what in the world novel elements of its models represent, but rather how they function in applications of the theory. Once one appreciates their functions it will become clear why these do not include representing elements of physical reality: novel quantum model elements such as wave functions do not represent what Bell [10] called beables of quantum theory.

By applying quantum models successfully, physicists have continued to make great progress without coming to agree on any interpretation_vF with its underlying physical ontology. That is what makes me optimistic that this pragmatist perspective is not just one among many, but the right way of viewing quantum theory. I am not a perspectival realist about the quantum world because I reject every interpretation_vF of quantum theory, though of course every concrete application of quantum theory is to something in the physical world. But I am a (non-perspectival) realist about quantum theory because that theory undoubtedly exists, and by focusing on its successful applications one can come to understand this theory and appreciate the ways in which it continues to contribute to scientific progress, now a century after its initial formulation.

A wave-function in a quantum model is used to represent a system's quantum state, and some quantum state assignments are true while others are false[1], so application of the model commits one to the existence of that quantum state, but not as a novel element of physical (or mental!) reality, observable or otherwise. Quantum states are just as real as the centre of mass of my desk and the current rate of inflation, neither of which is an element of physical reality, newly introduced by classical mechanics or macroeconomic theory (respectively).

However it may be represented mathematically, the function of a quantum state is not to represent an element of physical reality, but to offer advice to a hypothetical, physically situated, agent on the significance and credibility of descriptive claims about physical systems. If an agent is in this physical situation, they may benefit from this advice by adjusting their credences in significant claims to match the Born probabilities implied by assignment of the correct quantum state to the relevant physical system. An agent who is not in this situation may still use these probabilities to predict and in some cases to explain phenomena manifested by statistics that match them. An agent can apply quantum theory even to a possible world in which there are no agents by mentally "projecting" herself into the situation of an agent in that world.

The claims a quantum state offers advice about concern the values of physical magnitudes, such as the claim that an atom's internal energy is less than 30 electron volts, or that its *z*-spin is $\hbar/2$. They are not claims about the outcome of a measurement of that magnitude. But such a claim is significant only in a physical context of the sort that might permit its measurement: I call this a *decoherence environment*. Application of the Born rule to such a claim is licensed only if the quantum state of the system is robustly decohered by its environment so that the reduced state of the system stays very close to diagonal in a basis of eigenstates of operators associated with different values of that magnitude. Armed with the recommended credences, a situated agent has the information needed to make wise decisions about how to act. By projecting oneself into that situation, one can predict and understand the likely behaviour of similar systems in similar circumstances.

---
[1] Relative to what I call an agent-situation, as I will soon explain.

Born probabilities prescribe degrees of belief (credences) for *magnitude claims*, each of the form $M_s \varepsilon \Delta$. (T*he value of magnitude M on system* s *lies in Borel set* $\Delta \subseteq \Re$). Quantum no-go theorems (e.g. Kochen & Specker [11]) show that dynamical variables like (a component of) position and momentum can't all have precise real values at once. Each application of the Born rule must be restricted to magnitude claims that are all significant in that context, and therefore warrant some degree of belief—possibly zero. The physical process of measurement defines *some* such contexts, in which the outcome of measuring a magnitude M is specified by the truth of significant magnitude claims relevant to M.

Quantum models of the interaction of a system with its environment may help one decide whether a context is apt for application of the Born rule, and then to which magnitude claims. A magnitude claim $M_s \varepsilon \Delta$ may be assigned a Born probability only if it is meaningful to say it is true, or is false, relative to that context. This *licensing* use of the model is also advisory (cautionary): the unitary evolution of the system+environment quantum state in the model does not *represent* this as a physical process.

An epistemic agent can benefit from probabilistic advice on magnitude claims whose truth-value that agent is not in a position to access directly. Many such claims concern claims about the future whose truth-values are not determined by what happened in the agent's past. But an agent's physical situation may present other physical barriers to their knowledge of the world. Quantum theory's advice is useful because it is tailored to such situations. That is why quantum state assignments and Born probabilities are *relative*—not to a person or other agent, but to a physical situation they may be in. I call this an *agent-situation*, whether or not any agent is in that situation. An agent-situation is a physical viewpoint. It offers a meta-semantic and epistemic perspective on the physical world. Quantum states are assigned relative to a physical viewpoint also in Rovelli's relational quantum mechanics (RQM): [12] compares and contrasts my pragmatist perspective to RQM before briefly previewing section 4's argument for why in that perspective (the statistics of) our measurement outcomes provide objective evidence for quantum theory.

A system is not *in* a quantum state: its (perspectival) quantum state is *relative to* an agent-situation. So, the Born probability of a measurement outcome is also relative to an agent-situation. This is to be expected: the function of probability is to improve the epistemic situation of any agent whose physical situation presents a currently insuperable barrier to knowledge. Abstractly, both quantum state and Born probabilities are appearances from the viewpoint of an agent-situation. They are objective, not subjective or personal to any agent.

According to most textbooks, quantum theory predicts the Born probabilities of our observations: of alternative outcomes when we measure a magnitude. But those textbooks don't say what a measurement is, when it occurs, and whether there has to be someone observing its outcome. Now the second advisory function of a quantum state is to license the application of the Born rule to yield probabilities only of those mutually exclusive magnitude claims $M_s \varepsilon \Delta$ with enough content to be evaluable as true or as false. According to *inferentialist pragmatism* [13, 14], how much content a claim has is a function of the reliability of inferences to and from that claim. One should apply the Born rule to a magnitude claim only if many such inferences would be very reliable.

The quantum state in a model of environmental decoherence helps one to gauge inferential reliability. Each magnitude claim $M_s = m_j$ may be assigned a Born probability if the reduced quantum state of *S* in environment *E* is robustly diagonal in a "pointer basis" of eigenstates $|m_i>$ of $\widehat{M}$, the self-adjoint operator associated with M. The Born rule is applicable only to a magnitude claim $M_s = m_j$ on a system whose interaction with its environment closely meets that condition: I then call the physical context a *decoherence environment* for M [15].

When quantum theory is applied to model naturally-occurring physical interactions, decoherence environments occur naturally, with no agents making quantum measurements. Decoherence in such models is extremely rapid, robust, and practically irreversible. The Born rule is then applicable, whether or not an agent has arranged the conditions in which physical interactions produce a measurement outcome. A decoherence environment occurs in a space-time region that need not overlap the world-tube of an agent using it to perform a quantum measurement: it may be a process inside an evolving star.

Decoherence environments are also relative. What counts as a decoherence environment is specified by the environmental decoherence of a system's quantum state in a model. Since that quantum state is relative to agent-situation, what counts as a decoherence environment is also relative to agent-situation. This means that whether a measurement can be said to have an outcome is relative to agent-situation. In this sense *the outcome of a quantum measurement is perspectival.* The correlations between outcomes of spacelike separated measurements of different spin-components on each of two ions assigned an entangled EPR-Bohm state $1/\sqrt{2}(|\uparrow\downarrow> - |\downarrow\uparrow>)$ provides an example: see Fig. 1.

Suppose that Alice interacts with her ion in region *A*, measures *x*-spin up and reassigns the state of her ion to $1/\sqrt{2}(|\uparrow> + |\downarrow>)$ to successfully predict that her subsequent remeasurement will also yield spin up (as in [16]). By assigning state $1/\sqrt{2}(|\uparrow> - |\downarrow>)$ to Bob's ion, Alice correctly predicts probability ½ for each possible outcome of the *z*-spin measurement on his ion implemented by his interaction in region *B*. But since no quantum state describes an intrinsic property of the system to which it is assigned, Alice cannot infer that Bob's ion had *x*-spin down at any point in or on the light cone of region *B*. A model of environmental decoherence of Alice's ion encompassing region *A* as well as her own agent-situation in the future light cone of *A* permits her to take "*x*-spin up" to be a meaningful claim about her ion, and relative to that decoherence environment this truly states the outcome of her *x*-spin measurement.

Similarly, Bob measures *z*-spin up and reassigns the state of his ion to $|\uparrow>$ to successfully predict that his subsequent remeasurement will also yield spin up. A model of environmental decoherence of Bob's ion encompassing region *B* as well as his own agent-situation in the future light cone of B permits him to take "z-spin up" to be a meaningful claim about his ion, and relative to that decoherence environment this truly states the outcome of his z-spin measurement.

But no model of environmental decoherence of Bob's ion encompasses region B as well as Alice's agent-situation outside the future light cone of B, so Bob's measurement has no outcome relative to such an agent-situation. And no model of environmental decoherence of Alice's ion encompasses region A as well as Bob's agent-situation outside the future light cone of A, so Alice's measurement has no outcome relative to such an agent-situation.

On the other hand, there is a model of environmental decoherence encompassing both regions A and B as well as any agent-situation in the overlap of the future light cones of regions A and B. So relative to such an agent situation both Alice's and Bob's measurements may be said to have outcomes.

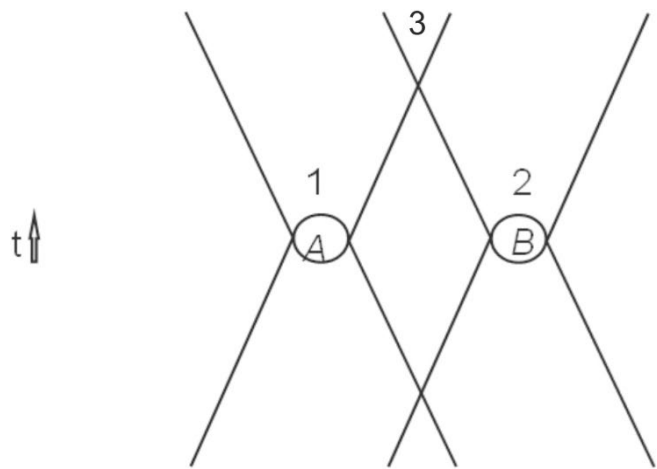

Fig. 1 Perspectival Outcomes in the EPR-Bohm state

This example shows how a measurement may have an outcome relative to one agent-situation but not relative to another. But note that the primary relativity here is not to an agent-situation but to a decoherence environment. Suppose that one alters exactly which spacetime location is the agent-situation relative to which the quantum state of each ion is assigned, within a region marked in the figure by the numeral '1', '2' respectively (in the future light cone of a region marked by the letter 'A', 'B' respectively, where the two light cones do not overlap). There is no significant alteration in the relevant decoherence environments, or in whether Alice's or Bob's measurement has an outcome relative to each altered agent-situation. The outcome of a quantum measurement is relative to a decoherence environment, and relative to an agent-situation only to the extent to which that decoherence environment is itself relative to agent-situation.

Relative to an agent-situation in region 3 (the overlap of the future light-cones of regions A and B), both Alice's and Bob's measurements may be said to have outcomes, because claims about their outcomes are both assessable at a decoherence environment for the joint measurement, relative to an agent-situation in region 3.[2] Only when Alice and Bob occupy agent-situations in region 3 should they agree that they have both performed quantum measurements with outcomes. Of course, physicists recording or reporting the result of an experimental realization of this scenario would indeed occupy such agent-situations.

## 4. Our measurement outcomes are perspectival but objective

One can describe thought-experiments in which quantum measurement outcomes must be relative to agent-situation, if the Born rule correctly predicts their probabilities. Because they extend a paradoxical scenario described by Eugene Wigner [17], these thought-experiments are called Extended Wigner's Friend scenarios (EWFS)s. In an EWFS each of several observers makes quantum measurements. One can argue that there are no absolute outcomes on which they could all agree: in some sense each outcome is relative to the observer. Some arguments are better than others: see [18]. But if the measurement outcomes whose observed statistics support the Born rule are similarly relative, then we seem to have little or no objective reason to accept quantum theory in the first place!

I will give a schematic account of one recent EWFS after first recalling the set-up of the scenario described by Wigner. He said that in this set-up the Friend agrees to make a measurement in her isolated laboratory and record the outcome, "collapsing" the quantum state of the system and the rest of her laboratory in accord with the eigenstate-eigenvalue link. But Wigner unitarily evolves the quantum state of the entire laboratory (including the Friend) and applies the eigenstate-eigenvalue link to conclude that the Friend's measurement had no outcome. The scenario is paradoxical because Friend and Wigner disagree, about the behaviour of the state and about whether a measurement occurred.

Here is the structure of an EWFS. Each of multiple friends is an agent confined to a (different) physically isolated laboratory in which they measure a magnitude on a quantum system and note the outcome. Outside that laboratory there is a "super-observer" agent: each has exquisite quantum control over everything in their friend's entire laboratory (including the friend). Because the isolation presents a physical barrier to each agent's knowledge of some other's outcomes, different agents assign some system distinct quantum states, each relative to the assigning agent's physical situation. Either a super-observer can (unitarily) "undo" their friend's quantum measurement, and then measure a different magnitude; or they can perform a specific precise measurement on their friend's entire lab that also erases all the friend's records of their own measurement outcome.

One extended Wigner's friend scenario considered in an interesting recent paper [19] presenting a no-go theorem for absolute measurement outcomes was described there like this. Inside their labs, the friends, Charlie and Debbie, each measure an observable on a different particle from a pair assigned an entangled state, producing their outcomes labelled $c$ and $d$, respectively. Outside their labs, "super-observers", Alice and Bob, each perform (ideally) space-like separated measurements labelled $x$ and $y$ on their particle of an incompatible observable after restoring the original entangled state of the pair by unitarily reversing Charlie's and Debbie's measurement interactions. Their outcomes are labelled *a*, *b* respectively. All quantum states evolve unitarily.

---

[2] Another way to think of this is to take the measurements by Alice and Bob to constitute a measurement of a single magnitude on the ion-pair system, represented by a tensor-product spin operator on its Hilbert space, with one of four possible outcomes relative to that decoherence environment.

In this and other EWFSs the hypothetical outcome frequencies don't match their quantum probabilities. Assume each friend and super-observer obtains some single outcome every time a measurement is repeated on a different system like that (on each "run"). Quantum theory predicts probabilistic correlations between pairs of these outcomes (in some EWFSs these correlations may be perfect). If each outcome is an absolute event that is relative to no-one and nothing else, then their relative frequencies in many runs must have a joint statistical distribution. But the probabilistic correlations between pairs of these outcomes predicted by quantum theory are incompatible with the existence of a joint probability distribution for all outcomes at once (such as *a*, *b*, *c*, *d* in this example). Given other plausible assumptions, if the quantum predictions are correct, then at least some of these measurement outcomes must be relative, not absolute.

Our evidence for quantum theory comes from measurements on physical systems. In an experiment when a measurement of the same type is repeated many times on similar systems assigned the same quantum state, we take there to be an observed relative frequency of outcomes that matches the probabilities quantum theory predicts. But if each measurement outcome is relative to the observer, then so are the experimental data themselves. If there are no absolute measurement outcomes for observers to agree on, then our experimental data provide no objective evidence for quantum theory. This is the new problem arising from recent no-go theorems in EWFSs.

This problem can be solved, in my pragmatist view. The solution comes in three stages:

1. Following quantum theory's advice on when to say a physical "measurement" interaction occurs.

2. Noting that, in our universe, external environmental interactions that can be neither eliminated nor shielded against are so pervasive that no EWFS will ever be realized.

3. Accepting that reports of outcomes of our measurements provide (strong) evidence for quantum theory that is objective in the only sense that matters for science.

To implement the first stage of the solution, recall how the question 'When is there a "measurement"?' is answered from this pragmatist perspective. One advisory function of a quantum state is to license the application of the Born rule to yield probabilities only of those mutually exclusive magnitude claims $M_s \varepsilon \Delta$ with enough content to be evaluable as true or as false. According to *inferentialist pragmatism* how much content a claim has is a function of the reliability of inferences to and from that claim. One should apply the Born rule to a magnitude claim only if many such inferences would be very reliable.

The quantum state in a model of environmental decoherence helps one to gauge inferential reliability. Each magnitude claim $M_s = m_j$ may be assigned a Born probability if the reduced quantum state of *S* in environment *E* is robustly diagonal in a "pointer basis" of eigenstates $|m_i>$ of $\widehat{M}$, the self-adjoint operator associated with M. The Born rule is applicable only to a magnitude claim $M_s = m_j$ on a system whose interaction with its environment closely meets that condition, in which case the physical context is a *decoherence environment* for M.

A physical "measurement" interaction (for M) occurs only in a decoherence environment for M. But the quantum state advising on whether this is a decoherence environment is relative to an agent-situation: being a decoherence environment for M is itself relative to agent-situation. So, whether a measurement of M occurs, and its outcome if it does, are both perspectival facts—facts relative to an agent-situation. That Friend has performed a measurement is a fact relative to her situation in the lab, but not relative to Wigner's situation outside.

Application of a quantum model of environmental decoherence requires assignment of a quantum state to a joint system $S+E$ of system and environment. Relative to a super-observer's agent-situation in an EWFS, $S$ includes the entire laboratory $L$ of their friend, and $E$ includes $L$'s external environment $X$. Relative to that friend's agent-situation, $S$ is a small quantum subsystem $Q$ of $L$, and $E$ is $Q$'s environment $I$ (everything else in $L$). If $L$ is physically isolated, there is no interaction between $X$ and $L$, so the quantum state assigned to $L$ by a super-observer in an agent-situation external to $L$ does not decohere.

The friend's measurement has no outcome relative to their super-observer's agent-situation. But there is an outcome relative to the friend's agent-situation, since the interaction between $Q$ and $I$ decoheres the quantum state of $Q$ relative to the friend's agent-situation (in the quantum model). *There is no absolute fact about whether the friend's measurement has an outcome*, because a decoherence environment relative to the friend's agent-situation is not a decoherence environment relative to the super-observer's.

Now we can move on to the second stage of the solution to the new problem posed by recent no-go theorems in EWFSs. External environmental interactions effectively prevent the physical isolation essential to an EWFS. Unwanted external environmental interactions are effectively impossible to eliminate or shield against, whether a physical "measurement" interaction occurs naturally, or scientists produce it to measure the value of a magnitude. Uncontrolled environmental decoherence is the main reason why it is so hard to build a reliable, large scale quantum computer.

No EWFS occurs naturally: realizing an EWFS deliberately would be enormously more difficult than building a reliable, large scale quantum computer. The "proof of principle" experiments performed to date [19, 20] come nowhere near realizing an EWFS, and there are reasons to remain extremely sceptical about the prospects of the ambitious research program aimed toward realizing an EWFS charted in a more recent publication [21]. Naturally occurring interactions modelled by quantum decoherence prevent the kind of isolation assumed in an EWFS: no experiment will ever realize a paradoxical EWFS.

This is important, because, if realized, a paradoxical EWFS would pose a serious threat to the objectivity of scientific knowledge based on quantum theory by undermining the assumption that a competently performed quantum measurement has an objective outcome. But our quantum measurements will never challenge that assumption.

Every type of competent quantum measurement that has ever been performed repeatedly in a laboratory experiment has generated outcome statistics matching Born probabilities, and we have no empirical reason to doubt that all future measurements will continue to do so. Each of these measurement outcomes states a perspectival fact—a fact relative to a physical context of assessment. But a basic principle of experimental methodology requires each experimenter to accept the indication of a reliable instrument, its record in a reliable recording device, and the report of a trustworthy colleague, as a fact relative to their own agent-situation.

By the time reports of the outcomes of measurements performed in a quantum experiment have been accepted as data, there is effectively only a single, shared, physical context of assessment relative to which each report states a fact. Even though this fact is relative to that context, it counts as objective in the sense that matters for scientific knowledge. To show this it is helpful to distinguish two ways in which such data might be claimed to be objective—transcendently and immanently.

Notions of objectivity can be applied much more widely than just to views of quantum mechanics since they concern truth and facts in general. Plain non-relative notions of truth and fact may be characterized by two principles.

*Truth*           A statement that $P$ is true if and only if $P$.

*Fact*            A statement states a fact if and only if it is true.

Corresponding relative notions may be characterized by these alternative principles:

*Relative Truth*  A statement that $P$ is true-relative to-$c$ if and only if $P$-relative to-$c$.

*Relative Fact*   A statement states a fact-relative to-$c$ if and only if it is true-relative to-$c$.

Here $c$ is a context at which a statement is assessed, not a context in which it is made: MacFarlane [22] distinguishes these contexts in developing his own account of relative truth. A variety of truth-relativism is associated with a class of statements and contexts for which no plain notion of truth and fact is applicable, but only notions of relative truth and relative fact.

It follows from the plain concepts of truth and fact that a statement that $P$ states a fact if and only if it is true; that is, if and only if $P$. In accordance with these concepts, a fact is a fact without regard to perspective. This irrelevance of perspective makes it natural (if overblown) to call a plain fact absolute or *transcendently objective*: it is absolute insofar as it is not relative to anything like a context or viewpoint, and transcendently objective as it is not limited by such things.

A perspectival statement is a statement assessable for truth only at a context. A relative fact is what is stated by a perspectival statement that is true as assessed at a context. A statement that is true relative to $c$ may not be true relative to a different context $c^*$: it may even be false relative to $c^*$, and so strongly relative. A context $c$ provides a semantic viewpoint on those perspectival statements assessable for truth at $c$. A perspectival statement assessed as true at context $c$ states a perspectival fact—a fact in the perspective offered by $c$: the set of such perspectival facts constitutes the perspective $P_c$ of $c$.

A relative fact in perspective $P_c$ may be a relative fact in the perspective $P_{c*}$ of *every* context $c^*$ at which it is assessed. That would not make this a plain fact because it is not simply true: the plain notion of truth is not applicable to a perspectival statement. It would rather make it a statement of (what may be called) an *immanently objective fact*.

After introducing the general notion of an immanently objective fact, one can now apply it to the specific case of reports of quantum measurement outcomes. In this case, a context of assessment for one or more magnitude claims is a decoherence environment at which those claims are sufficiently meaningful to be assessed for truth or falsity. Such decoherence environments occur when one or more magnitudes are measured. This application of the notion of an immanently objective fact is important because statistical summaries of the outcomes of such quantum measurements are the data scientists treat as the evidence against which they test predictions of the Born rule. When these statistics are judged to match predicted Born probabilities in a wide range of cases while never significantly deviating from them, that is convincing evidence warranting acceptance of quantum theory.

Science depends on immanently objective facts here because these are what all sincere inquirers can come to agree on as assessed at essentially one common decoherence environment, despite any inconsequential differences in their agent-situations. The scenario depicted earlier involving the EPR-Bohm state provides an example of this. All agent-situations in region 3 (the overlap of the future light cones of regions $A$, $B$) effectively share a single context of assessment, both for a magnitude claim about the value of the $x$-spin of Alice's ion and for a magnitude claim about the value of the $z$-spin of Bob's ion. The joint outcome of Alice's and Bob's measurements can be recorded only in region 3, and any subsequent observation of that record will also occur in region 3. It follows that there is effectively only a single context of assessment for this pair of magnitude claims shared by all agents able to observe or access records of both Alice's and Bob's measurement outcomes, no matter how their agent-situations differ.

More generally, if there were no immanently objective facts then there would be nothing scientists would be justified in acknowledging as *data*, capable of confirming or refuting scientific knowledge claims. Science depends on immanent objectivity but not transcendent objectivity. Because scientists and other observers in our world come to share basically the same context of assessment, all our quantum measurement outcomes are immanently objective (relative) facts. But the perspectival character of these facts implies no relativism of scientific knowledge based on quantum theory, because our changing physical situations always privilege a single, shared perspective, in which reports of the outcomes of our quantum measurements are immanently objective facts. These provide us with all the evidence we need to warrant acceptance of quantum theory. Of course, because science is fallible, we should bear in mind that subsequent evidence might warrant a different attitude toward any specific quantum theory, or even toward the whole framework of quantum theories.

# References

[1]     Healey, R.A.:  Pragmatism Works: Essays on Quantum Theory, Science and Metaphysics. Oxford University Press, Oxford (forthcoming)
[2]     Healey, R.A. The Quantum Revolution in Philosophy. Oxford University Press, Oxford (2017).
[3]     Jacoby, F.: "Perspectivism in Science", *Internet Encyclopedia of Philosophy*. Accessed 27th October 2024
[4]     Perspectivism. Wikipedia. Accessed 27th October 2024
[5]     Cabello, A. Interpretations of quantum theory: A map of madness. In: Lombardi, O., Fortin, S., Holik, F., and López, C. (eds.) What is Quantum Information? Cambridge, Cambridge University Press (2017), 138-143: arXiv:1509.04711 [quant-ph]
[6]     Mermin, N.D. Quantum mechanics: Fixing the shifty split. Physics Today 65 No. 7, 8–10 (2012).
[7]     Wallace, D. On the Plurality of Quantum Theories: Quantum Theory as a Framework, and its Implications for the Quantum Measurement Problem. In French, S. and Saatsi, J. (eds.) Scientific Realism and the Quantum. Oxford University Press, Oxford (2020)
[8]     Wittgenstein, L.  Philosophical Investigations, tr. Anscombe, G.E.M. Basil Blackwell, Oxford (1968)
[9]     Van Fraassen, B.C. Quantum Mechanics: an Empiricist View . Oxford, Oxford University Press (1991), p.9.
[10]    Bell, J.S. Speakable and Unspeakable in Quantum Mechanics, revised edition. Cambridge University Press, Cambridge (2004)
[11]    Kochen, S. and Specker, E.P. The problem of hidden variables in quantum mechanics. Journal of Mathematics and Mechanics 17, 59-87 (1967)
[12]    Healey, R.A. Securing the objectivity of relative facts in the quantum world. Foundations of Physics 52: 88, 1-20 (2022)
[13]    Brandom, R. Making it Explicit. Harvard University Press, Cambridge, Mass. (1994)
[14]    Brandom, R. Articulating Reasons. Harvard University Press, Cambridge, Mass. (2000)
[15]    Healey, R.A. Scientific objectivity and its limits. British Journal for the Philosophy of Science 75 (3), 639-62 (2024)
[16]    Wang, P. *et al*. Significant loophole-free test of Kochen-Specker contextuality using two species of atomic ions. Science Advances 8, eabk1660 (2022)
[17]    Wigner, E. P. Remarks on the mind-body question. In: Good, I.J., (ed) The Scientist Speculates. Heinemann, London (1962).
[18]    Healey, R., Quantum theory and the limits of objectivity.  Foundations of Physics 48, 1568–89 (2018)
[19]    Bong, K.-W. *et al*. A strong no-go theorem on the Wigner's friend paradox. Nature Physics 16, December, 1199–1205 (2020) https://www.nature.com/articles/s41567-020-0990-x
[20]    Proietti, M. *et al*. Experimental test of local observer independence. Science Advances 5 (9), (2019) https://www.science.org/doi/10.1126/sciadv.aaw9832
[21]    Wiseman, H. M, Cavalcanti, E. G. and Rieffel, E. G. A "thoughtful" Local Friendliness no-go theorem: a prospective experiment with new assumptions to suit. Quantum 7, 1112 (2023) https://quantum-journal.org/papers/q-2023-09-14-1112/.
[22]    MacFarlane, J. Assessment Sensitivity. Clarendon Press, Oxford. (2014)